\documentclass[fleqn,usenatbib]{mnras}

\usepackage{newtxtext,newtxmath}

\usepackage[T1]{fontenc}
\usepackage[normalem]{ulem} 
\usepackage{bm}

\DeclareRobustCommand{\VAN}[3]{#2}
\let\VANthebibliography\thebibliography
\def\thebibliography{\DeclareRobustCommand{\VAN}[3]{##3}\VANthebibliography}

\usepackage{graphicx}	
\usepackage{amsmath}	

\usepackage{hyperref} 
\hypersetup{linkcolor=red,citecolor=blue,filecolor=cyan,urlcolor=blue}

\newcommand\Teff{T_\mathrm{eff}}

\title[Accretion onto a new-born neutron star]{Post-Supernova Accretion of Light Elements onto a New-Born Neutron Star and NS 1987A}

\author[N. de J. Baz-Pérez et al.]{
Natalia de Jes\'us Baz-P\'erez,$^{1}$\thanks{E-mail: nbaez@astro.unam.mx}
Dany Page,$^{1}$
Simon Guichandut,$^{2}$
Martin Nava-Callejas,$^{1,3}$
\newauthor
Yuri Cavecchi,$^{4}$
Andrew Cumming$^{2}$
\\
$^{1}$Instituto de Astronom\'ia, Universidad Nacional Aut\'onoma de M\'exico, Ciudad de  M\'exico, CDMX 04510, Mexico
\\
$^{2}$Department of Physics and Trottier Space Institute, McGill University, 3600 rue University, Montreal, QC, H3A 2T8, Canada
\\
$^{3}$Institut d'Astronomie et d'Astrophysique, Université Libre de Bruxelles, 1050 Bruxelles, Belgique
\\
$^{4}$Departament de Fis\'{i}ca, EEBE, Universitat Polit\`ecnica de Catalunya, Av. Eduard Maristany 16, 08019 Barcelona, Spain
}

\date{Accepted XXX. Received YYY; in original form ZZZ}

\pubyear{2025}

\begin{document}

\maketitle

\begin{abstract}
We model early accretion of light elements, He, C, and O, onto a new-born neutron star using the public stellar evolution code  \texttt{MESA}, simulating what may happen during the first few years of its life.
We find that, under the appropriate conditions, significant amounts of these elements can be accreted  up to densities of $10^9$ g cm$^{-3}$  without triggering a nuclear explosion that would convert them into heavy elements.
These results help to understand observations that favor light elements in the atmospheres of young cooling neutron stars, as the one found in the supernova remnant Cassiopeia A, and also add support to the recent indications for the presence of a neutron star, NS 1987A, in the remnant of SN 1987A.
\end{abstract}


\section{Introduction} \label{sec:intro}

The supernova SN 1987A was one of the major events in the history of modern astronomy \citep{Arnett:1989aa}, the first one visible to the naked eye for several centuries \citep{1987IAUC.4316....1K}, observed in all wavelengths and with a couple of dozen of neutrinos detected, emitted by the just formed proto-neutron star
\citep{Bionta:1987aa,Hirata:1987aa}.
The detected neutrinos implied that a neutron star was born \citep{Burrows:1986aa,1987ApJ...318L..63B} but further detection of this object has so far eluded all searches 
(see, e.g., \citealt{2007AIPC..937..134M,Esposito+2018} and \citealt{Alp:2018aa}), a fact that
led to the speculation that the proto-neutron star may have collapsed into a black-hole \citep{1992ComAp..16..153B}.
The first serious hint for the presence of a neutron star in the remnant of SN 1987A was obtained by the ALMA observatory through the detection, at sub-millimeter wavelengths, of a small blob of warm dust (``the blob'' hereafter), close to the center of the explosion, whose higher temperature above the surroundings would be due to heating by a compact object \citep{Cigan:2019aa}.
It was then argued by \citet{Page:2020aa} that the most likely source of heating is the thermal radiation from a hot young neutron star, dubbed ``NS 1987A'', that naturally emits $\sim 10^{35}$ erg s$^{-1}$ at an age of a few decades, close to the estimated luminosity irradiating the blob.
Alternate mechanisms for heating of the blob as accretion onto a compact object, either a black hole or a neutron star, cannot be ruled out as well as heating from a forming pulsar wind nebula, but both possibilities require fine tuning of parameters.
Analysis of X-ray observations of SN 1987A do indicate the presence of a pulsar wind nebula \citep{Greco:2021aa,Greco:2022aa}, but other interpretations have been proposed \citep{Alp:2021aa}.
Strong further evidence for the presence of a neutron star has been obtained from JWST observations of SN 1987A which revealed narrow infrared emission lines of argon and sulfur that were interpreted as due to the ionizing radiation from a young compact object, with a slight preference in favor of a neutron star thermal radiation over a pulsar wind nebula as the source of the radiation \citep{Fransson:2024aa,Larsson:2025aa}
\footnote{Another set of observations with the JWST \protect\citep{Bouchet:2024aa} did not find any evidence in favor of the presence of the neutron star, but it was not covering the wavelength range in which the argon and sulfur lines are seen.}.

The cooling of a young neutron star is totally driven by neutrino emission (particularly, in the outer layers of the crust, by the plasmon decay process, \citealt{Adams:1963oz,Nomoto:1981aa,Beznogov:2020aa}) while the surface photon emission is only a tracer of the interior evolution as its contribution to the energy losses is negligible until the star is at least several tens of thousands of years old (see, e.g., \citealt{Yakovlev:2004aa,2006NuPhA.777..497P,Potekhin:2015aa}).
However, the star's effective temperature, $\Teff$, and its surface thermal luminosity, are determined by how easily heat from the hot ($\sim 10^9$ K) interior is allowed to flow toward the surface.
This heat flow is most sensitive to the layers where electrons dominate the transport and is controlled by their scattering off the nuclei \citep{Gudmundsson:1983aa}.
If these upper layers of the star, its {\it envelope}, are composed of light elements (such as H, He, C, O, with low electric charge $Z$) this heat flow will be stronger, and result in higher $\Teff$, than in the presence of heavy elements (such as Fe, with larger $Z$) as shown by
\citet{Chabrier:1997aa} and \citet{Potekhin:1997aa}.
If the energy source powering the blob identified by ALMA is the thermal radiation from NS 1987A, this would favor the presence of a thick layer of light elements on the surface of the neutron star:
\citet{Page:2020aa} showed that in the case of an envelope made of heavy elements the present surface luminosity of NS 1987A would be slightly below $10^{35}$ erg s$^{-1}$ 
while the presence of a sufficiently thick layer of light elements can raise this luminosity to a few times $10^{35}$ erg s$^{-1}$, in better agreement with the
estimated luminosity of the blob, $1-3 \times 10^{35}$ erg s$^{-1}$ \citep{Cigan:2019aa}.

Theory predicts that, being formed at extremely high temperatures, the upper layers of a new-born neutron star should be made of $^{56}$Fe and similar iron peak nuclei \citep{Lattimer:1991ab,Haensel:2007aa} with no light elements present.
The presence of light elements in the envelope of NS 1987A would then imply that these have been deposited by accretion of material after the neutron star was formed. Strong fall-back \citep{Colgate:1971aa} is found in many core collapse supernova simulations and, moreover, hypercritical accretion \citep{Blondin:1986aa}, happening within a few hours after the explosion, likely occurred in the case of SN 1987A as initially proposed by \citet{Chevalier:1989aa}.
Such gigantic early mass accretion rates, however, generate very high temperatures in the compressed layers ($\sim 10^{10}$ K) and still result in the formation of an envelope made of heavy elements.
We consider that deposition of light elements at the surface of a new-born neutron star would be due to accretion of matter which failed to be ejected by the explosion but that occurs on time scales of months to years.
Such late accretion is similar to accretion in a binary system and is limited to rates lower than or approaching the Eddington rate, $\dot{M}_\mathrm{Edd} \sim 10^{18}$ g s$^{-1}$
\citep{Frank:2002aa}, itself resulting in a luminosity $L_\mathrm{Edd} \sim 10^{38}$ erg s$^{-1}$.
The bolometric light-curve of SN 1987A \citep{Suntzeff:1990aa,Branch:2017} displays a luminosity above $L_\mathrm{Edd}$ during the first three years, which dropped below 10\% of this value during the fourth year. 
This gives a window of a few years during which strong accretion could have occurred and deposited a thick layer of light elements on the surface of NS 1987A.

It is our purpose in the present paper to study this accretion phase onto a new-born, hot, neutron star at a rate close to the Eddington limit.
Many accreting neutron stars in low-mass X-ray binaries frequently exhibits X-ray bursts during which explosive burning of the accreted H/He results in massive production of heavy elements, even well beyond the iron peak: such unstable burning, if occurring during accretion onto a new-born neutron star, would convert most of the accreted light elements into heavy ones. 
Consequently, of particular interest is not only the possibility of accretion, but more importantly the conditions under which such accretion of light elements can avoid unstable nuclear burning and how thick a layer of light elements can thus remain at the surface of a new-born neutron star.

We  describe in \S~\ref{sec:setup} our setup while our results can be found in \S~\ref{sec:results}.
In \S~\ref{sec:evidences} we discuss arguments in favor of the presence of light elements at the surface of young neutron stars.
Finally, in \S~\ref{sec:concl} we offer a discussion and our conclusions.

\section{Modeling accretion onto a new-born neutron star with \texttt{MESA} } \label{sec:setup} 

For this purpose, we employ the public stellar evolution code \texttt{MESA}, version 15140
\citep{Paxton:2011aa,Paxton:2013aa,Paxton:2015aa,Paxton:2018aa,Paxton:2019aa,Jermyn:2023aa}.
This code is able to model accretion onto a neutron star with chosen chemical composition of the matter and mass accretion rate
\citep{Paxton:2015aa,Zamfir:2014aa,Meisel:2018aa,Guichandut:2023aa,Song:2024aa,Zhen:2025aa,Nava-Callejas:2025ab,Nava-Callejas:2025aa}.
For the envelope we use the default equation of state and opacity of \texttt{MESA} that can handle densities up to $10^{11}$ g cm$^{-3}$ \citep{Paxton:2011aa,Paxton:2013aa,Jermyn:2023aa}.

We will model a neutron star envelope initially made of $^{56}$Fe with a maximum density at its base $\rho_b = 10^{10}$ or $10^{11}$ g cm$^{-3}$.
To generate the initial $T-\rho$ profile we start with the model provided in the \texttt{MESA} test suite \texttt{make\_env}
that has a bottom density of $\rho_b = 6 \times 10^{6}$ g cm$^{-3}$ with an inner boundary luminosity $L_b = 1.13 \times 10^{34}$ erg s$^{-1}$ coming from the star's deeper layer.
In a first step we turn on accretion of pure $^{56}$Fe, assuming a star of mass $M=1.4 \, M_\odot$ and radius $R=10$ km,  until the model base density has reached $\rho_b = 10^{10}$ g cm$^{-3}$.
In a second step, without accretion, we generate a series of models adjusting slightly $L_b$ and letting each one evolve in time until it reaches a stationary state.
Notice that, once the stationary state has been reached, each of these models, being composed exclusively of $^{56}$Fe with given $M$ and $R$, is uniquely determined by just two parameters, $\rho_b$ and $L_b$, and gives us as output its bottom temperature $T_b$ as well as its surface temperature $T_s$.
These models are replicas of the stationary envelope models described by \cite{Gudmundsson:1983aa} and we verified that their bottom temperature $T_b$ and surface temperature $T_s$ agree with the ``$T_b - T_s$'' relationship of Eq. (32) of these authors.
We repeated the process accreting now various amounts of light elements, with nuclear burning turned off, and then evolving them without accretion to a stationary state. In these cases, we found good agreement  with the ``$T_b - T_s$'' relationship of the stationary accreted envelope models described by \cite{Potekhin:1997aa}.
These preliminary results give us confidence that \texttt{MESA} is able to reliably describe thick stationary neutron star envelopes with various chemical compositions.

For our present study we finally generated a stationary model with $\rho_b = 10^{11}$ g cm$^{-3}$ and  $L_b = 10^{35}$ erg s$^{-1}$  composed of pure $^{56}$Fe.
This base luminosity is the luminosity predicted to flow from the interior into the envelope of a new born neutron star at an age of a few years, as described by \cite{Beznogov:2020aa}, which is the range of ages during which we propose strong accretion may have occurred on NS 1987A.
Small variations of $L_b$ are possible, either due to its weak time dependence during the first few years of the star's evolution or under variations in the assumed stellar mass and/or radius \citep{Page:2020aa}, so we will also consider some cases with $L_b = \frac{1}{2} \times 10^{35}$ erg s$^{-1}$and
$2 \times 10^{35}$ erg s$^{-1}$.

Having this initial basic stationary envelope model, we can simulate accretion of light elements to investigate how thick a layer of these can be deposited onto the surface of a new-born neutron star without being converted to heavy elements in an explosion due to unstable burning.
We consider accretion of $^{4}$He, $^{12}$C, and $^{16}$O in various proportions, as these three nuclides are the most likely light ones to be accreted considering the ``onion skin structure'' of the interior of massive stars that explode in core collapse supernovae \citep{2013sse..book.....K}.
3D core collapse simulations, e.g., \citet{2016ARNPS..66..341J}, as well as observations of young supernova remnants such as Cassiopeia A \citep{2010ApJ...725.2038D}, show that extensive mixing is occurring during the explosion and thus it cannot be excluded that H may also be accreted.
In this first work, we do not consider accretion of H:
this alleviates the necessity of employing large nuclear networks for an adequate simulation of the rp-process in the presence of hydrogen \citep{1981ApJS...45..389W} which render the numerical simulations much more demanding.
We will address this issue of the possible presence of H in a future work.

\begin{figure*}
\centerline{\includegraphics[width=0.99\linewidth]{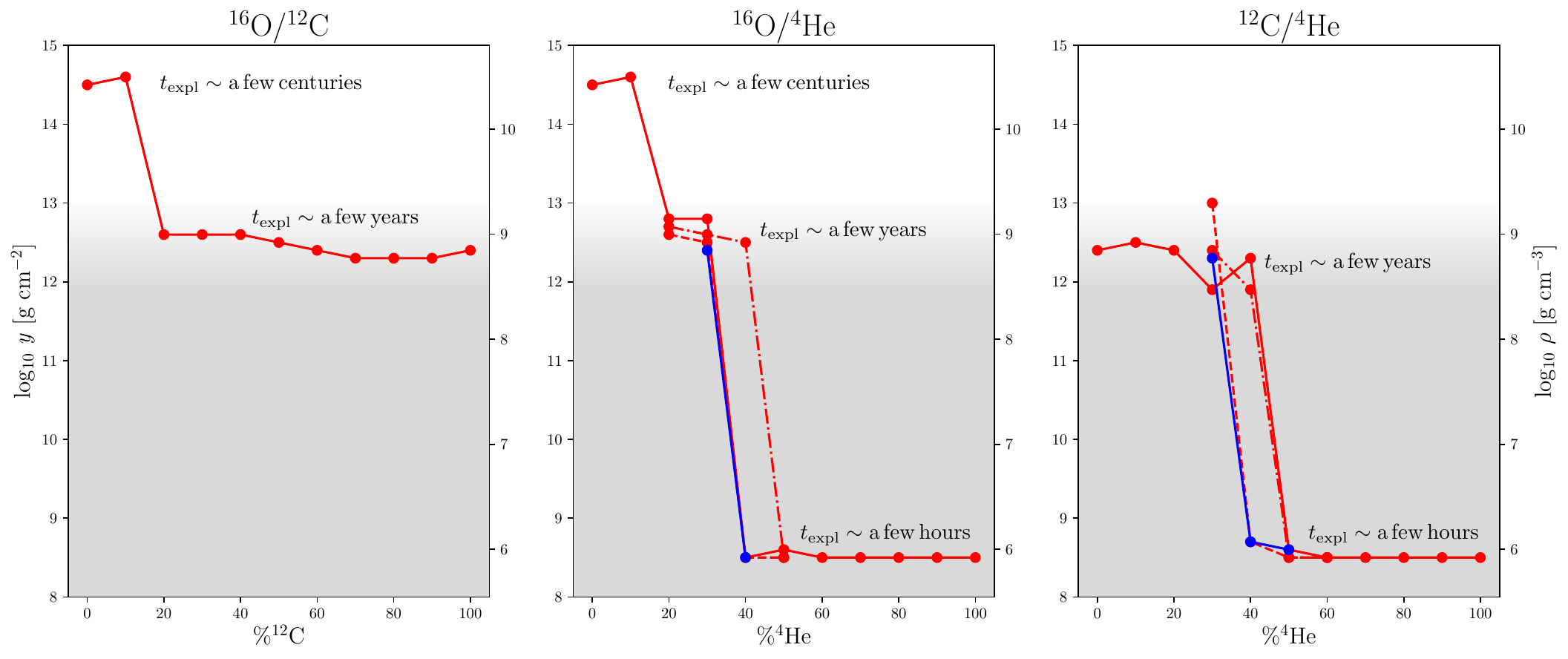}}
\caption{Summary of our results: column density $y$ (and density $\rho$) reached by the accreted matter before an explosion occurs, as a function of its chemical composition for our three families of models with mixtures of  $^{16}$O/$^{12}$C, $^{16}$O/$^{4}$He, and $^{12}$C/$^{4}$He.
Red lines correspond to $\dot{M} = 10^{-8} \, M_\odot$ yr$^{-1}$ and blue lines to $\dot{M} = 2.6 \times 10^{-8} \, M_\odot$ yr$^{-1} \simeq \dot{M}_\mathrm{Edd}$.
We apply three base luminosities: 
$L_b = 10^{35}$ erg s$^{-1}$, continuous lines,
$\frac{1}{2} \times 10^{35}$ erg s$^{-1}$, dashed lines, and
$2 \times 10^{35}$ erg s$^{-1}$, dash-dotted lines.
Order of magnitude explosion times, $t_\mathrm{expl}$, are indicated: at a few centuries
explosions are triggered by unstable $^{16}$O burning, at a few years by unstable $^{12}$C burning and at a few hours by $^{4}$He.
The gray backgrounds indicate the range of values accessible in up to three years of accretion at the Eddington rate $\dot{m}_\mathrm{Edd} \sim 10^5$ g cm$^{-2}$ s$^{-1}$, the maximum possible heavy accretion time in the case of SN 1987A.
Accretion that stops before reaching the displayed explosion depth will result in a thick light element envelope, while if it reaches the explosion depth most of the accreted matter will be converted into heavy elements.
}
\label{fig1}
\end{figure*}

We employ the \texttt{MESA} provided nuclear network \texttt{approx21}\footnote{This network is described on the cococubed site,
  \href{https://cococubed.com/code_pages/burn_helium.shtml}{https://cococubed.com/code\_pages/burn\_helium.shtml}.}.
We fix the mass accretion rate at $\dot{M} = 10^{-8} M_\odot$ yr$^{-1}$, i.e.,
$\dot{m} = \dot{M}/(4 \pi R^2) \sim 3\times 10^{4}$ g cm$^{-2}$ s$^{-1}$, which is roughly one third of the Eddington rate, and follow the evolution of the envelope until an explosion occurs: such explosion will convert most of the light elements into heavy ones. 
To result in a thick light element envelope, early accretion should have occurred during a time shorter than this explosion time.
For completeness, we will also consider a few cases at the Eddington rate.

\section{Results } \label{sec:results} 

\begin{figure*}
\centerline{\includegraphics[width=0.99\linewidth]{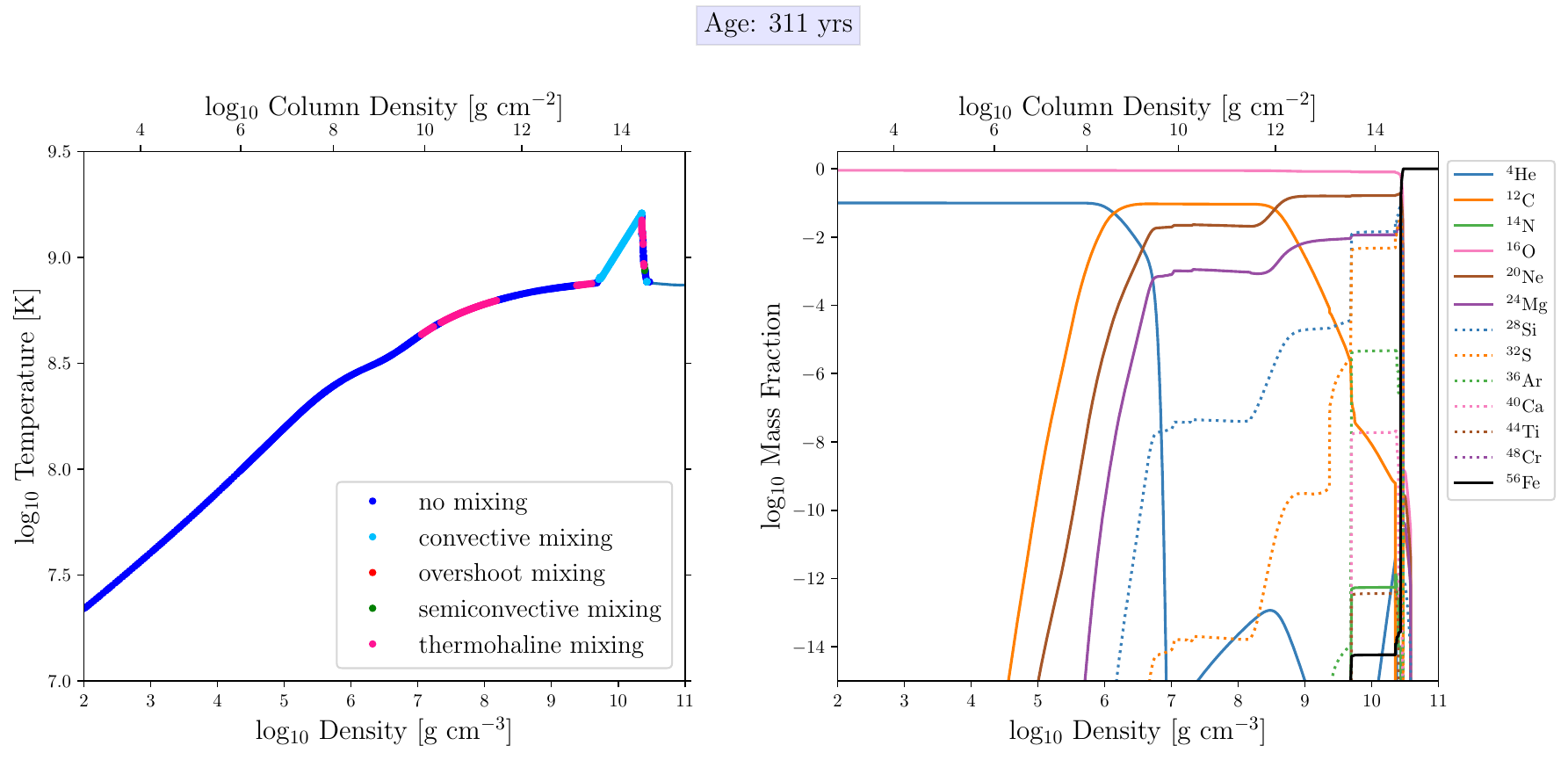}}
\caption{Example of a model where both $^{4}$He and $^{12}$C are exhausted before they reach their critical explosion densities and an explosion is triggered by unstable $^{16}$O burning.
The left panel shows the temperature profile after $\sim 311$ years of accretion of 10\% $^{4}$He immersed in 90\% $^{16}$O at a rate $\dot{M} = 10^{-8} \, M_\odot$ yr$^{-1}$.
Typical of all our simulations, this model exemplifies the high temperatures present in these envelopes, above $7\times 10^8$ K at $\rho \sim 10^{11}$ g cm$^{-3}$, as a result of the high flux coming from the hot young neutron star interior.
In the right panel one can appreciate the depletion and then exhaustion of $^{4}$He at column depths between $10^8$ to $10^{10}$ g cm$^{-2}$ resulting in the appearance of $^{12}$C.
Notice also the appearance of $^{20}$Ne, and $^{24}$Mg, as well as traces of heavier elements from succesive $(\alpha, \gamma)$ captures on $^{16}$O, $^{20}$Ne, $^{24}$Mg, $\dots$.
Initiation of burning of $^{12}$C through $^{12}$C --$^{12}$C fusion is seen at $y \sim 10^{12}$ g cm$^{-2}$, resulting in the production of $^{20}$Ne and $^{24}$Mg, and traces of $^{28}$Si and $^{32}$S, rapidly leading to its depletion.
Deeper, at $y \sim 3 \times 10^{13}$ g cm$^{-2}$, $^{16}$O burning from $^{16}$O --$^{16}$O (and some $^{12}$C --$^{16}$O) fusions is initiated and becomes unstable, triggering an explosion seen as a temperature peak in the left panel.
}
\label{fig2}
\end{figure*}

We ran three families of models depending on the accreted matter chemical composition, starting with pure $^{16}$O, $^{12}$C, or $^{4}$He, to which we then add a second nucleus, chosen among the same three, in various fractions.
Models are run with \texttt{MESA} until an explosion is triggered.
We summarize our results in Figure \ref{fig1} and first discuss the basic scenario with $L_b = 10^{35}$ erg s$^{-1}$ and $\dot{M} = 10^{-8} \, M_\odot$ yr$^{-1}$
(continuous red lines in the figure).

In the cases of accretion of pure $^{16}$O we find it takes of the order of centuries to reach an explosion: it happens when the $^{16}$O reach an explosion density $\rho_\mathrm{expl}^\mathrm{O} \sim 3 \times 10^{10}$  g cm$^{-3}$.
In the case of pure $^{12}$C, an explosion occurs when the accreted matter reaches a depth $y_\mathrm{expl}^\mathrm{C} \sim 3 \times 10^{12}$ g cm$^{-2}$, at a density $\rho_\mathrm{expl}^\mathrm{C} \sim 10^9$  g cm$^{-3}$, after 2.6 years of accretion.
For pure $^{4}$He it takes only a few hours to reach explosive conditions occurring when matter reaches a depth $y_\mathrm{expl}^\mathrm{He} \sim 3 \times 10^{8}$ g cm$^{-2}$, at a density $\rho_\mathrm{expl}^\mathrm{He} \sim 10^6$  g cm$^{-3}$.
Note that these critical densities for unstable burning of $^{4}$He and $^{12}$C are similar to the ones found in models of accreting low-mass binaries \citep{Strohmayer:2003aa,2021ASSL..461..209G}.

When adding $^{12}$C to $^{16}$O we see two timescales for explosion (left panel of Figure \ref{fig1}).
At low $^{12}$C concentration, 10\%, an explosion occurs after a few centuries while at higher concentrations it occurs after a few year: in the low concentration case $^{12}$C has stably burned into $^{20}$Ne and $^{24}$Mg and became exhausted
before reaching $\rho_\mathrm{expl}^\mathrm{C}$ 
so that the explosion is due to unstable burning of $^{16}$O
while at higher concentration it survived till $\rho_\mathrm{expl}^\mathrm{C}$ and triggered an explosion when reaching it within a few years.

Similarly, adding $^{4}$He to $^{12}$C we again see a pattern with two timescales (right panel of Figure \ref{fig1}).
As long as the $^{4}$He mass fraction is less than 40\%, $^{4}$He burns stably into $^{12}$C and is exhausted before reaching $\rho_\mathrm{expl}^\mathrm{He}$. An explosion occurs when $^{12}$C reaches  $\rho_\mathrm{expl}^\mathrm{C}$ after a few years.
At higher concentrations, $^{4}$He survives up to $\rho_\mathrm{expl}^\mathrm{He}$ resulting in an explosion within a few hours.
One such model is described in more details below in Sec \ref{sec:Case1}.

When adding $^{4}$He to $^{16}$O (central panel of Figure \ref{fig1}) three timescales appear.
While the $^{4}$He fraction is at most 10\% we find almost the same pattern as when adding $^{12}$C to the $^{16}$O.
In these cases, $^{4}$He burns stably into $^{12}$C which itself burns stably into heavier elements, both before reaching their respective $\rho_\mathrm{expl}$, so that the explosion occurs when $^{16}$O reaches $\rho_\mathrm{expl}^\mathrm{O}$, requiring a centuries long accretion phase.
Such a case is illustrated in Figure~\ref{fig2}.
At higher concentrations, but below 40\%, $^{4}$He still burns stably into $^{12}$C. The latter is produced in sufficiently large amounts to survive up to $\rho_\mathrm{expl}^\mathrm{C}$, resulting in an explosion after a few years of accretion.
Once the $^{4}$He fraction reaches 40\% and beyond, the explosion time drops to a few hours and corresponds to an $^{4}$He triggered explosion due to the $3\alpha$ reaction, the same pattern as found in the case of adding $^{4}$He to $^{12}$C.

We finally explore the sensitivity of these results to changes in the two model parameters $L_b$ and $\dot{M}$. 
Increasing $L_b$ will increase the temperature and thus the burning rates, possibly exhausting nuclei before they reach their exploding density.
On the other hand, increasing $\dot{M}$ results in pushing nuclei faster toward their exploding density. This can either avoid exhaustion or accelerate it as increasing $\dot{M}$ also results in higher temperatures in the envelope.
We can consider these effects only briefly here, leaving a detailed assessment to future work, and restrict ourselves to the transition of stable/unstable burning of $^{4}$He.
We first explore the case of variations in $L_b$ by either a factor $\frac{1}{2}$ or $2$: these are shown as dashed and dash-dotted (red) lines in Figure \ref{fig1}.
In the case of the mixture $^{16}$O/$^{4}$He a small decrease in $L_b$ does not change the stable/unstable threshold while a small increase in $L_b$ does raise the threshold by 10\%.
In the case of the mixture $^{12}$C/$^{4}$He a small decrease in $L_b$ does reduce the stable/unstable threshold by 10\% while a small increase in $L_b$ does not.
When considering a change in the mass accretion rate, increasing $\dot{M}$ to $\dot{M}_\mathrm{Edd}$ (blue lines in the figure) does not result in a change in the critical $^{4}$He fraction that result in explosion when it is mixed with $^{16}$O.
However, when mixed with $^{12}$C we find that the threshold concentration is {\it decreased} when  $\dot{M}$ is increased:
the model with 40\% $^{4}$He that was stable at $\dot{M} = 10^{-8} \, M_\odot$ yr$^{-1}$ becomes unstable at higher $\dot{M} = 2.6 \times 10^{-8} \, M_\odot$ yr$^{-1}$.
In this particular case the increasing speed, which pushes nuclei $^{4}$He faster to their explosion density, wins over the increasing temperature, which burns $^{4}$He faster and depletes it.
In all cases we find that the discriminant between stable and unstable burning is the possible exhaustion of He before it reaches its ignition depth.

We now discuss in more details two particular cases. 
The first one describes the burning of $^{4}$He and the explosion it triggers, representative of all short explosion times, while the second one examines a case of $^{4}$He exhaustion and the subsequent explosive burning of the produced $^{12}$C, all immersed in a background of $^{16}$O, representative of the few years long explosion time-scales.

\subsection{A first case study: accretion of carbon and helium} \label{sec:Case1} 

\begin{figure*}
\centerline{\includegraphics[width=0.95\linewidth]{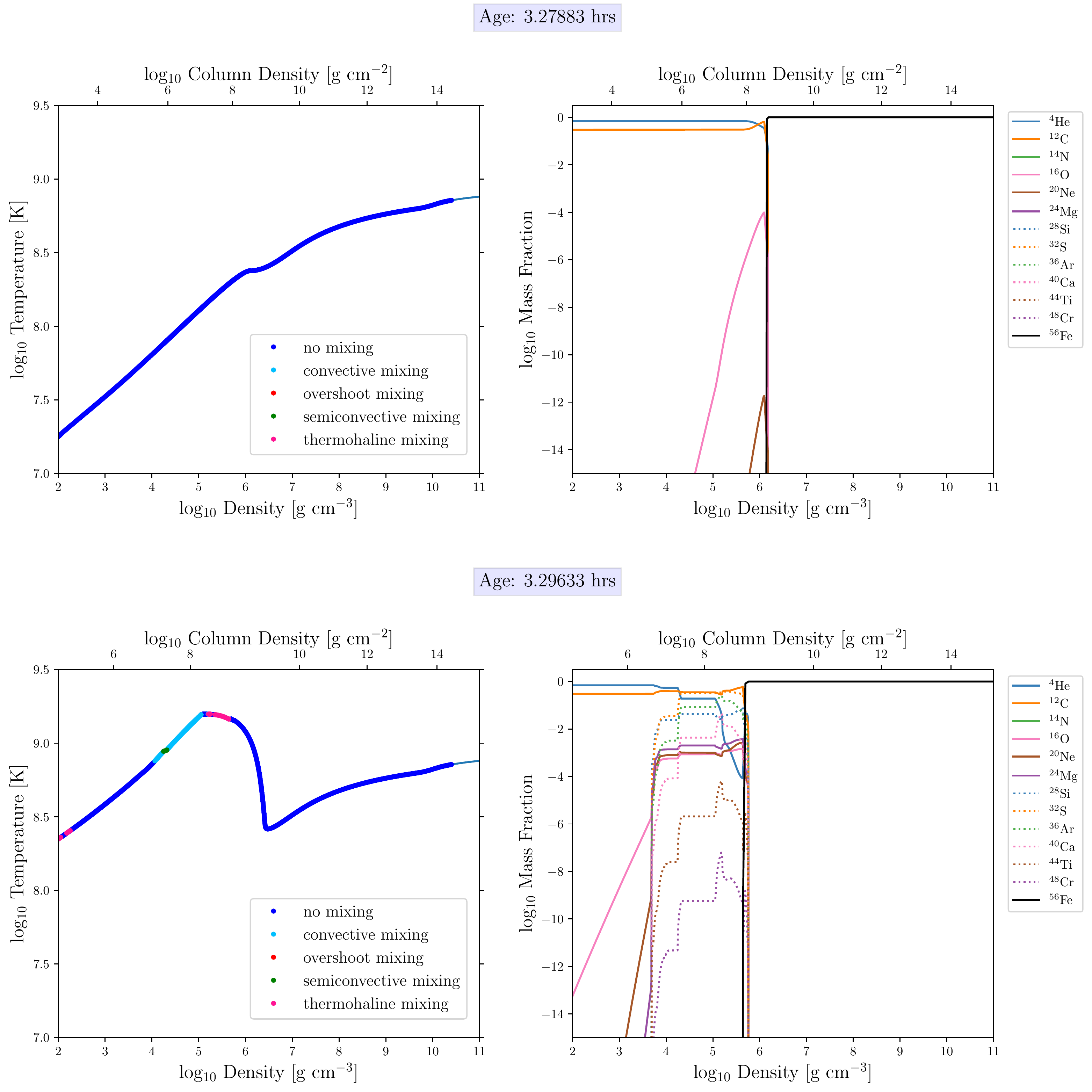}}
\caption{Model accreting 30\% $^{12}$C and 70\% $^{4}$He at a rate $\dot{M} = 10^{-8} \, M_\odot$ yr$^{-1}$.
The upper panel is taken at a time previous to the explosion and the lower panel illustrates the explosion itself.
In each panel the left frames present the $T-\rho$ profile of the envelope, at the indicated time, with coloring depicting the hydrodynamic state of the various layers.
The right frames present the composition of matter at the same times.
\\
An animated version of this figure will be available with the on-line version of the paper (MNRAS) or 
from the arXiv abstract page either as an "Ancillary File" or by downloading the "Tex Source" package.
}
\label{fig3}
\end{figure*}

In Figure \ref{fig3} we present details of the model that accreted 30\% $^{12}$C and 70\% $^{4}$He.
This model explodes after three and a quarter hours and the upper panel describes it 
a few minutes before the explosion, while the lower panel is at a time when the explosion has fully developed.
The explosion is triggered by $^{4}$He burning through the $3\alpha$ reaction becoming unstable and occurs when the accreted matter reaches a column depth $\sim 3\times 10^8$ g cm$^{-2}$, at a density slightly above $10^6$ g cm$^{-3}$.
In the upper panel, before the explosion is triggered, one notices that at the highest densities reached by the accreted matter most of the $^{4}$He has been stably converted into $^{12}$C, and some further $(\alpha, \gamma)$ captures have produced small amounts of $^{16}$O and $^{20}$Ne.
Notice also the very high temperature, about $8\times 10^8$ K, at the bottom of the envelope, a typical value for the outer crust of a new-born neutron star \citep{Beznogov:2020aa}
and much higher than typical crust temperatures, less than a few times  $10^8$ K,  in accreting neutron stars
(see \citealt{2022ApJ...933..216P} for temperatures in the crust of the hottest observed accreting neutron star).
Once matter has been pushed to slightly higher densities, with higher temperatures, burning through the $3\alpha$ reaction becomes unstable and the lower panel depicts the envelope once the explosion is well developed, with a typical temperature peak higher than $10^9$ K, as seen in the left frame.
Two convective zones are present (marked in light blue on the temperature profile of the left frame) resulting in flat mass fraction profiles in the right frame.
Heavy nuclei are being produced now, by series of $(\alpha, \gamma)$ captures, which are already reaching $^{48}$Cr in the figure.
This explosive burning will eventually lead to nuclear statistical equilibrium, i.e., iron peak nuclei, but the details of the generation of nuclear statistical equilibrium are beyond the reach of the simplified \texttt{approx21} nuclear network.
Notice, moreover, that the explosive burning propagates toward lower densities but stalls when it reaches densities between $10^3-10^4$ g cm$^{-3}$ 
\citep{Hanawa:1982ab,Weinberg:2006aa},
leaving the outermost part of the envelope untouched.
Most of these outer layers may, however, be ejected in case the burst reaches Eddington luminosity (\citealt{Weinberg:2006aa}, \citealt{Guichandut:2023aa}).
Another noticeable feature is the density reached by the accreted matter, which decreases due to fluid expansion at such high temperatures $\sim 10^9$ K, but will increase again later after the explosion once matter cools down.

\subsection{A second case study: accretion of oxygen and helium} \label{sec:Case2} 

\begin{figure*}
\centerline{\includegraphics[width=0.95\linewidth]{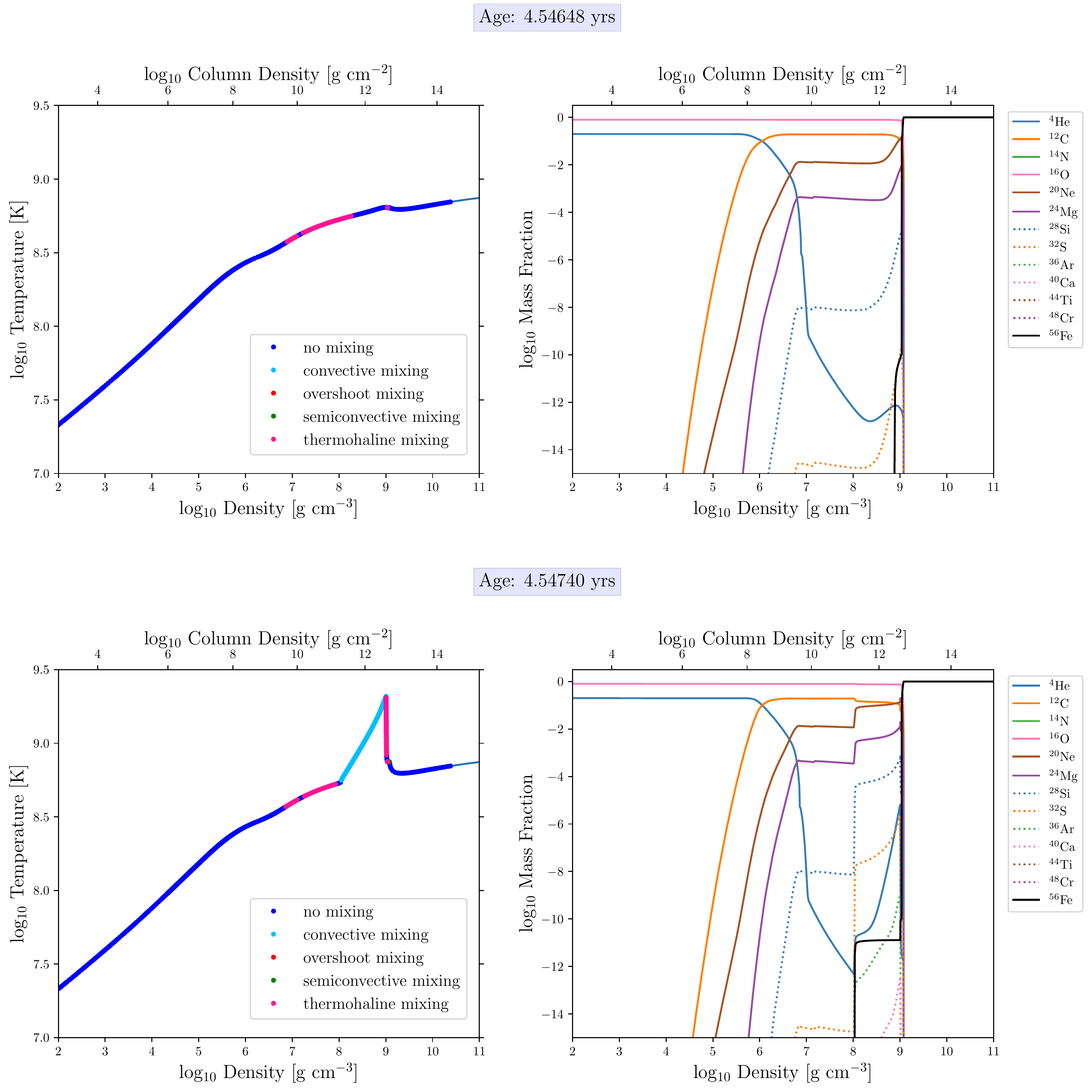}}
\caption{Model accreting 80\% $^{16}$O and 20\% $^{4}$He at a rate $\dot{M} = 10^{-8} \, M_\odot$ yr$^{-1}$.
The panels and their frames are analogous to the ones of Figure \ref{fig3}.
\\
An animated version of this figure will be available with the on-line version of the paper (MNRAS) or 
from the arXiv abstract page either as an "Ancillary File" or by downloading the "Tex Source" package.
}
\label{fig4}
\end{figure*}

In Figure \ref{fig4} we present details of a second model that accreted 80\% $^{16}$O and 20\% $^{4}$He.
In contradistinction to the previous case, in this model helium burning does not become unstable and an explosion is triggered much later, at an age of 4.55 year, by the burning of $^{12}C$.
Helium is seen to become depleted above a column density of $\sim 10^{8}$ g cm$^{-2}$ ($\rho \sim 10^{6}$ g cm$^{-3}$) 
and is exhausted before it reaches a column density  of $\sim 10^{10}$ g cm$^{-2}$ ($\rho \sim 10^{7}$ g cm$^{-3}$).
This helium has been essentially burned into $^{12}$C and $^{16}$O, and smaller amounts of $^{20}$Ne, $^{24}$Mg, and traces of heavier nuclei, due to $(\alpha, \gamma)$ captures.
We see, in the upper panel, that approaching $y \sim 10^{12}$ g cm$^{-2}$ ($\rho \sim 10^{9}$ g cm$^{-3}$) the carbon and oxygen fractions begin to decrease noticeably due to their fusion and their products, $^{20}$Ne, $^{24}$Mg, and $^{28}$Si, are increasing.
Once pushed to slightly higher densities, these fusion reactions, $^{12}$C$+$$^{12}$C and $^{12}$C$+$$^{16}$O, become unstable and trigger an explosion which is depicted in the lower panel of the same figure.
As in the previous case study, this explosion will lead to nuclear statistical equilibrium of iron peak nuclei, but it is very likely that matter at low densities will escape processing leaving some superficial layer of light elements, as is seen in models of explosive $^{12}$C burning in superbursts
\citep{2011ApJ...743..189K, 2012ApJ...752..150K}.

\section{Evidence for the presence of light elements at the surface of young neutron stars}
\label{sec:evidences} 

There is some limited evidence for the presence of light elements at the surface of young isolated cooling neutron stars, deduced from several characteristics of their observed soft X-ray thermal spectra.


On one hand there are several observations of absorption lines.
For example, two (\citealt{Sanwal:2002aa}, possibly three or even four, \citealt{Bignami:2003aa}) absorption lines have been observed in the cases of 1E 1207.44-5209 but are most naturally interpreted as electron cyclotron line and its harmonic(s) in a magnetic field of $~ \sim 10^{11}$ G (\citealt{Sanwal:2002aa}, \citealt{Bignami:2003aa}, \citealt{Pavlov:2009a}, \citealt{Halpern_2011a}).
While other interpretations have been proposed as, e.g., due to atomic transitions of He$^+$ \citep{Sanwal:2002aa}, molecular hydrogen \citep{Turbiner:2004aa} or even O and Ne \citep{Hailey:2002aa}, all in the presence of a strong magnetic field, they do not provide unambiguous identification of the surface chemical composition but would imply the presence of light elements.
Other examples of observed absorption lines are the ``X-ray Dim Isolated Neutron Stars'' (XDINS, see, e.g. \citealt{Turolla:2009ab} for a review), unfortunately with the same ambiguous results, although future observations 
are likely to shed more light in this case \citep{Mancini-Pires:2023aa,Kurpas:2024aa}. 


Even in the absence of absorption lines, the shape of the emitted spectrum strongly depends on the chemical composition of the atmosphere and spectral fit with different atmosphere models 
result in very different inferred effective temperatures and sizes of the emitting area (see, e.g., \citealt{Romani:1987aa}).
In the case where the distance $D$ to the star is known, its luminosity can be deduced in two ways: either directly from the observed flux $F$ once corrected for interstellar absorption, as $L=4\pi D^2 F$, and, alternatively, from the spectrally inferred temperature as $L = 4\pi R_\mathrm{eff}^2 \sigma_\mathrm{SB} T_\mathrm{eff}^4$,
where $R_\mathrm{eff}$ is an effective emission radius and $\sigma_\mathrm{SB}$ the Stefan-Boltzmann constant.
Comparison of the two gives $R_\mathrm{eff}$ which should be of the order of 10-15 km, assuming the whole stellar surface is emitting at an approximately uniform temperature, which would be the case in the absence of a strong magnetic field.
The ``Central Compact Objects'' (CCOs, \citealt{Pavlov:2002aa}, \citealt{De-Luca:2017aa}) are thought to be such weakly magnetized neutron stars and this approach has provided evidence for the presence of C at the surface of the neutron stars CXO J232327.8+584842 (``Cas A'', \citealt{Ho:2009aa}), XMMU J173203.3-344518 \citep{Klochkov:2013aa}, and  CXOU J181852.0-150213 \citep{Klochkov:2016aa}, as well as possibly CXOU J160103.1-513353 \citep{Doroshenko:2018aa}.
The presence of a weak magnetic field in these stars is inferred from the non detection of a modulation of the X-ray emission, which would be a sign of a strong magnetic field that induces surface temperature non-uniformities \citep{Page:1995aa}.
However, since these stars are located within young X-ray bright supernova remnants, limits on this amplitude modulation are unfortunately not very strong and, moreover, unfavorable geometry may also hide strong surface temperature inhomogeneities.
If such were the case, then a two-component blackbody spectrum can also give a good fit to the thermal X-ray spectrum of these sources, weakening the inference of the presence of C at their surface \citep{Alford:2023aa}.
Such an unfavorable geometry would, however, have to be present in four cases, making it quite improbable.


In the case of stars with a strong magnetic field, this $R_\mathrm{eff}$ argument is not so convincing as only a reduced fraction of the surface may be hot enough to be detected, particularly in the presence of a strong toroidal component
\citep{2006A&A...457..937G,2006A&A...451.1009P}.
Moreover, atmosphere models in the presence of a strong magnetic field are not as reliable as the non-magnetic ones.
The best models are for atmospheres composed of H and such models
are consistent with the observed spectrum, and give acceptable values of $R_\mathrm{eff}$, in the case of the Vela pulsar \citep{1996rftu.proc..173P,2001ApJ...552L.129P} 
while the situation in not as clear in the cases of other young pulsars (see, e.g., \citealt{2009ASSL..357..181Z}).
In the case of several middle-aged pulsars (the "Three Musketeers" PSR B0656+14, PSR B1055-52 and Geminga, as well as PSR J1741-2054) magnetized H atmospheres are not able to describe the observed spectra (see, e.g., \citealt{Abramkin:2025a} and references therein).


Unfortunately, evidence for the presence of light elements at the surface of a neutron star does not give us any information about the actual thickness of the light element layer beneath the surface, because the amount of accreted matter needed to result in an optical thickness of unity, $\sim 1 \, \mathrm{g \, cm}^{-2}$, is extremely small, $\sim 4\pi R^2 \times 1\, \mathrm{g \, cm}^{-2} \sim 10^{13}$ g $\sim 10^{-20}$ $M_\odot$. Moreover, in the absence of strong accretion, as in the case of isolated cooling neutron stars, while sedimentation will rapidly bring the lightest element to the surface \citep{Alcock:1980aa}, contributing to thicken this light element layer, diffusive burning will slowly consume H and/or He \citep{2003ApJ...585..464C,2010ApJ...723..719C,2019MNRAS.484..974W}. Furthermore, as it is always seen in simulations of short X-ray bursts  - and observed in our simulations as well - the nuclear burning does not reach $\rho\leq 10^4$ g cm$^{-3}$, implying that even in these cases we still have sufficient amounts of light elements at the surface to result in a thin envelope enriched in low-$Z$ material. Therefore, it is very difficult to estimate the thickness of this layer from observations.


Where such light elements at the surface of a neutron star come from is not clear.
Accretion from interstellar medium needs $L_\mathrm{PSR} \ll 10^{30}$ erg s$^{-1}$ \citep{Ostriker:1970aa} and seems to be excluded for all known pulsars and the three CCOs with measured $P$ and $\dot{P}$. 
However, since only a very small amount of accreted matter is needed to result in a light element atmosphere, one cannot completely exclude this possibility.
In this paper we presented arguments that indicate how heavy post-supernova accretion is a viable alternative and the inferred luminosity of the putative NS 1987A also argues in this direction.

\section{Discussion and Conclusions} \label{sec:concl} 

The observed bolometric light curve of SN 1987A \citep{Suntzeff:1990aa,Branch:2017} implies that the strong mass accretion rates we are considering, $\dot{m} = (0.3-1) \times 10^5$ g cm$^{-2}$ s$^{-1}$, could have lasted at most 3 years within which the accreted matter could have reached a depth of $y_L \sim 10^{13}$ g cm$^{-2}$, corresponding to a density above $\rho_L \sim 10^{9}$ g cm$^{-3}$.
After an initial relaxation phase of a few months, the luminosity coming out of the crust of a new born neutron star is, within a factor of two, around $10^{35}$ erg s$^{-1}$ \citep{Page:2020aa}.
Under these conditions, our results, summarized in Figure \ref{fig1}, show that a thick layer of $^{16}$O and/or $^{12}$C can be accreted as long as the accreted matter does not contain too much $^{4}$He, less than about 30-40\%.
When the mass fraction of $^{4}$He is larger than 30-40\%, $^{4}$He burning becomes unstable and leads to a thermonuclear explosion when He reaches densities 
$\sim 10^6$ g cm$^{-2}$, converting most of the accreted matter to iron peak nuclei and leaving only a thin layer, up to densities $10^{(3-4)}$ g cm$^{-3}$, of unburnt light elements which, moreover, may be ejected if the burst reaches Eddington luminosity \citep{Weinberg:2006aa,Guichandut:2023aa}.

However, if $\dot{M}$ could be higher than $\dot{M}_\mathrm{Edd}$ or $L_b$ larger than implied by the results of \citet{Beznogov:2020aa} then 
$^{4}$He burning could be stable \citep{Zamfir:2014aa}. $^4$He would turn into $^{16}$O/$^{12}$C before reaching densities of $\sim 10^6$ g cm$^{-3}$ and this would result, after a few years of accretion, in an envelope made of a thick layer of $^{16}$O/$^{12}$C beneath a thin layer of $^{4}$He.
It is not clear under which conditions $\dot{M}$ could be larger than $\dot{M}_\mathrm{Edd}$, but a base luminosity above our applied $L_b$ could be possible.

In the case of neutron stars accreting in low-mass X-ray binaries it has been found that an energy source, called ``shallow heating'', whose nature is still unknown, should be present in the outer envelope of the stars \citep{Brown:2009aa,Wijnands:2017a,Nava-Callejas:2025aa}. 
That such an energy source is acting in a new-born neutron star is speculative 
and moreover its effect in such hot envelopes as we are studying may not be significant.
Thus, if present, it may or may not increase $L_b$ and result in stable burning of $^{4}$He.
In such a case a thick envelope of light elements could be generated even in cases where our study predicts otherwise.

On the other hand, $^{12}$C burning may become unstable well before our modeling infers, because the commonly employed $^{12}$C --$^{12}$C fusion cross-section is possibly underestimated in the astrophysically relevant energy range.
This energy range is below 2 MeV and no experimental data are available at such low energies (see \citealt{Nan:2025aa} for a recent discussion) so that this cross-section is an extrapolation that could be a strong underestimate in the case of the presence of resonances \citep{Cooper:2009aa}.

It is worth mentioning that observations of accreting neutron stars in low-mass X-ray binaries show that bursts appear to be quenched when the mass accretion rate is above $\sim 0.2-0.3 \, \dot{M}_\mathrm{Edd}$  \citep{2003A&A...405.1033C,2008ApJS..179..360G,2021ASSL..461..209G}
while 1D numerical modeling still produces bursts up to much higher accretion rates \citep{Bildsten:1998aa,2007ApJ...665.1311H,Fisker_2007,Nava2024}.
It has been suggested that this discrepancy between theory and observations could be resolved considering in more details the geometry of the accretion flow \citep{Cavecchi:2020aa,Cavecchi:2025aa}.
If the same effects apply to SN 1987A, our numerical simulations may have resulted in bursts in cases that actually should not have exploded so that some of the cases reported as explosive in Figure \ref{fig1} may actually be stable.
However, if the fallback matter does not form a disc as in the low-mass X-ray binaries discussed by \citet{Cavecchi:2025aa}, those geometrical arguments may not be applicable in our case.

When comparing with Figures 2 and 3 in \cite{Page:2020aa}, we see that a light element envelope reaching $\rho_b \sim 10^9$ g cm$^{-3}$ implies a surface luminosity for a new-born neutron star that is within the lower range of $1-3 \times 10^{35}$ erg s$^{-1}$ needed to power the blob where NS 1987A is supposedly hiding in the debris of SN 1987A, adding credit to the claims of the existence of this neutron star \citep{Cigan:2019aa,Fransson:2024aa,Larsson:2025aa}.

DATA and Software:

No new data were generated or analysed in support of this research


\section*{Data Availability}

The \texttt{MESA} release 15140 is available from \href{https://docs.mesastar.org}{https://docs.mesastar.org}.
The \texttt{inlist} employed to perform the \texttt{MESA} runs is available with the on-line version of the paper.

This work made use of the following software:
\texttt{PYTHON} (\href{https://www.python.org/}{https://www.python.org/}),
\texttt{NUMPY} \citep{Harris:2020aa},
\texttt{MATPLOTLIB }\citep{Hunter:2007aa},
and \texttt{Mesa\_Reader} 
(\href{https://github.com/wmwolf/py_mesa_reader}{https://github.com/wmwolf/py\_mesa\_reader})


\section*{Acknowledgements}
N.J.B.-P. and M.N.-C. acknowledge support from a fellowship of Conahcyt.
Their and D.P.'s work is also supported by a UNAM-DGAPA grant PAPIIT-IN114424.
A.C. and S.G. were supported by Discovery Grant RGPIN-2023-03620 from the 
Natural Sciences and Engineering Research Council of Canada (NSERC)
Y.C. acknowledges support from the grant RYC2021-032718-I, financed by MCIN/AEI/10.13039/501100011033 and the European Union NextGenerationEU/PRTR.
M.N.-C. acknowledges support by the Fonds de la Recherche Scientifique-FNRS under Grant No IISN 4.4502.19.
Finally, the authors thank G. Pavlov for comments and suggestions about \S~\ref{sec:evidences}.
\bibliographystyle{mnras}


\input{NS1987A_Accretion_MNRAS.bbl}


\end{document}